\documentclass[12pt,preprint]{aastex}
\shorttitle{$AKARI$ NIR/MIR spectroscopy of APM 08279+5255}
\shortauthors{S. Oyabu et al.}
\title{$AKARI$ near- and mid-infrared spectroscopy of APM 08279+5255
  at $z=3.91$$^\dagger$} 
\author{S. Oyabu\altaffilmark{1}, K. Kawara\altaffilmark{2},
  Y. Tsuzuki\altaffilmark{3}, Y. Matsuoka\altaffilmark{2,*}, 
  H. Sameshima\altaffilmark{2}, N. Asami\altaffilmark{2} and
  Y. Ohyama\altaffilmark{4}}
\altaffiltext{$\dagger$}{Based on observations with AKARI, a JAXA project with the participation of ESA.}
\altaffiltext{1}{Institute of Space and Astronautical Science, Japan
  Aerospace Exploration Agency, 3-1-1 Yosinodai, Sagamihara, Kanagawa
229-8510, Japan}
\altaffiltext{2}{Institute of Astronomy, 
  University of Tokyo, 2-21-1 Osawa, Mitaka, Tokyo
181-0015, Japan}
\altaffiltext{3}{System Integration Service Department, Redfox Inc.,
  3-3-11, Kita-Aoyama, Minato-ku, Tokyo 107-0061, Japan}
\altaffiltext{4}{Academia Sinica, Institute of Astronomy and
  Astrophysics, Taiwan}
\altaffiltext{*}{Research Fellow of
the Japan Society for the Promotion of Science}
\email{oyabu@ir.isas.jaxa.jp}

\begin{document}

\begin{abstract}

We present rest-frame optical/near-infrared spectra of the gravitationally lensed quasar 
APM 08279+5255 at $z=3.91$ that has been taken using
the Infrared Camera (IRC) onboard the $AKARI$ infrared satellite.
The observed continuum consists of two components; a power-law 
component dominating optical wavelengths which is the direct light from 
the central source and thermal emission dominating near-infrared 
wavelengths which is attributed to the emission from hot dust in the 
circumnuclear region. 
The thermal emission well represents optically thick emission by hot
dust at $T \sim 1300$K with $\tau_{2\mu\mathrm{m}} > 2$ and apparent
mass, $M_{\mathrm{hot}}>10 \mathrm{M}_{\sun}$. 
Thus, our observations directly
detected the optically thick region of hot dust in APM 08279+5255.
HI recombination lines of H$\alpha$(0.656\micron), 
Pa$\alpha$(1.875\micron), and Pa$\beta$(1.282\micron) are clearly detected at 
3.2, 6.3, and 9.3 \micron. Simulations with the 
photoionization models suggest that APM 08279+5255 has BLR(Broad Line Region) 
clouds characterized by $\log n_{H} \sim 12- 14$ for the gas density, 
$\log U \sim -2 - -6$ for the ionization parameter, 
and $E(B-V)\sim 0.3-0.6$ for the broad line region. Thus, optically
thick emission of hot dust support an idea on non-spherical
distribution of dust near the central source, consistent with the
Active Galactic Nuclei model with the dust torus.
The temperature of hot dust and flux ratios of these HI lines are similar to 
those observed in low-redshift quasars.
There are significant time-variations in the HI lines, which are probably 
caused by variations in the brightness of the central source.
\end{abstract}
\maketitle

\section{Introduction}

High-redshift quasars provide direct probes of the distant early
universe where galaxies and quasars formed. Over the past decade, 
many high-redshift quasars have been found in large-area surveys 
such as the Sloan Digital Sky Survey \citep{york00} and Two Degree Field (2dF) 
QSO Redshift Survey \citep{boyle00}. Nonetheless, no convincing evidences 
for evolution in quasar spectra have been found, which is in remarked 
contrast to spectra of galaxies. For example, 
the flux ratio \ion{Fe}{2}/\ion{Mg}{2} in quasars is expected to be small 
at high-redshift because the Fe-enrichment is delayed relative to 
the $\alpha$-element enrichment \citep[i.e.][]{yoshii98}. However, 
\ion{Fe}{2}/\ion{Mg}{2} does not change from low-redshift to high-redshift 
\citep{wills85,tsuzuki06,kawara96,iwamuro04,kurk07,barth03}. Other UV
emission lines such as \ion{N}{5} and 
\ion{C}{4} show no evidence for evolution either \citep{dietrich03,nagao06}.
 
Dust locating at the circumnuclear region is heated by the central engine,
and produce near-infrared radiation. Such hot dust is common in Active
Galactic Nuclei (AGNs) and quasars at low redshift and also observed 
in high-redshift quasars \citep{oyabu01,hines06,jiang06}. 
Dust at high redshift is presumably produced as a result of supernova 
explosions, while dust production would be dominated by mass-loss of late-type 
stars at low-redshift. If so, hot dust at high-redshift would be different 
from that at low-redshift. Unfortunately, hot dust in the
high-redshift has been mostly studied 
based on broad-band photometry \citep{oyabu01,hines06,jiang06}, and no
systematic evolution has been reported.  
Spectroscopic studies would provide new insight into evolution and origin of 
hot dust in the circumnuclear region of quasars.

$AKARI$, which was launched in 2006 February, is the first Japanese 
satellite dedicated to infrared observations \citep{murakami07}.
It has a 68.5 cm diameter telescope LHe-cooled to 5.4 K. In addition to its 
major mission that is to perform all-sky survey at six bands in the mid- and 
far-infrared, $AKARI$ has carried out pointed observations for deep surveys 
of selected area and systematic observations of important objects. 

We have performed the near- and mid-infrared spectroscopy of the quasar
APM 08279+5255 at $z=3.91$ using Infrared Camera
\citep[IRC:][]{onaka07} onboard $AKARI$. These observations were
obtained as part of  $AKARI$ Open Time program ``IRC NIR
Spectroscopy of High-Redshift Quasars''. 

APM 08279+5255 was discovered during a survey for high Galactic
latitude carbon stars \citep{irwin98}. The IR flux densities at 25, 60, and
100 \micron are 0.23, 0.51, and 0.95 Jy, respectively, 
in the $IRAS$ Faint Source
Catalog \citep{moshir92}. This is apparently one of the most luminous 
objects with $5 \times 10^{15} L_{\sun}$ at $z=3.91$ \citep{downes99}.
The ground-based imagery observations in the optical \citep{ibata99} and 
near- and mid-infrared \citep{egami00}, revealed that 
APM 08279+5255 is gravitationally lensed with a
magnification of $\sim100$.
Using the Infrared Spectrograph on the {\it Spitzer Space
Telescope}, \citet{soifer04} detected broad Pa$\alpha$ and Pa$\beta$
HI recombination lines as well as a strong, red continuum
in the rest-frame wavelength range 1-7\micron.

Throughout this paper, $\Omega=0.3,\Lambda=0.7$, and $h=0.7$ are assumed for 
cosmology parameters. 

\section{Observation and Data Reduction}
\label{sec:obs}

Spectroscopy was performed on APM 08279+5255 using the 
IRC \citep{onaka07,ohyama07} onboard the $AKARI$ satellite 
\citep{murakami07} on 2006 October and 2007 April.
The observations are summarized in Table \ref{tab:obslog}.

We used two IRC's channels, NIR and MIR-S. 
Each covers different wavelengths from the
near-infrared to mid-infrared. The NIR channel uses a 512 $\times$ 412 InSb
array, whereas the MIR-S employs a 256 $\times$ 256 Si:As array.
The NIR and MIR-S channels share the same field of views and observe 
simultaneously. 

The Astronomical Observational Template 04 (AOT04) designed for spectroscopy 
was used. AOT04 replaces the imaging filters by transmission-type 
dispersers on the filter wheels to take near- and mid-infrared spectra. 
For 2006 October observations, the NIR Prism (NP) was set to cover
the wavelength of 1.8-5.5\micron~
with a spectral resolving power of $R\sim$19 at 3.5\micron. The target
was put on the center of the detector array without a slit.
In the MIR-S channel, two grisms,
SG1(4.6-9.2\micron;R$\sim$53) and SG2(7.2-13.4\micron;R$\sim$50), were set 
and slitless spectroscopy was made.
For 2007 April observations, the NIR grism (NG) was set to cover the wavelength
of 2.5 - 5\micron~with the resolution R $\sim120$. The 1\arcmin~$\times$ 1\arcmin~slit was used to avoid the confusion of spectra.

The data were processed through the IRC-dedicated data reduction
package, IRC\_SPECRED Ver. 20080528 \citep{ohyama07}. First, dark subtraction, 
linearity correction, flat correction, and various image anomaly corrections 
were performed. Multiple exposures were then coadded. After performing 
wavelength- and flux-scaling, the spectra of the object was extracted.
Apertures for the spectrum extraction were
5 pixels, namely,  (NP, SG1, SG2, NG) = (7.5\arcsec, 12.5\arcsec,
12.5\arcsec, 7.5\arcsec).
Aperture correction was also performed at the end. 

We found large uncertainties associated with the SG1 and SG2 calibration 
at $\lambda_{\mathrm{obs}}>
8.3$\micron~and  $\lambda_{\mathrm{obs}}> 12.0$\micron, respectively. This 
was caused by the second order light affecting the
response curves of both grisms. Thus, in this work, we only use the
wavelength at $\lambda_{\mathrm{obs}}<
8.3$\micron~and  $\lambda_{\mathrm{obs}}< 12.0$\micron~for 
the SG1 and SG2 spectra, respectively.

\begin{deluxetable}{llrrrr}
    \tabletypesize{\scriptsize}
    \tablecaption{Observational log and modes (AOT04)\label{tab:obslog}}
    \tablehead{
        \colhead{Obs. Date} & \colhead{Pointing ID} &
        \colhead{Channel/Disperser} 
        & \colhead{Wavelength} & \colhead{Resolving power}
        & \colhead{Extraction aperture} \\
        (UT) & & & (\micron) &  & 
      }
      \startdata
        2006-10-19 & 3120025.001 & NIR/NP & 1.8-5.5 & 19 at
        3.5\micron & 7.5\arcsec\\
                               & & MIR-S/SG1 & 4.6-9.2 & 53 at
                               6.6\micron & 12.5\arcsec\\
                               & & MIR-S/SG2 & 7.2-13.4 & 50 at
                               10.6\micron & 12.5\arcsec\\
        2007-04-16 & 3120050.001 & NIR/NG & 2.5-5.0 & 120 at
        3.6\micron & 7.5\arcsec\\
        \enddata
\end{deluxetable}

\begin{figure*}
    \plotone{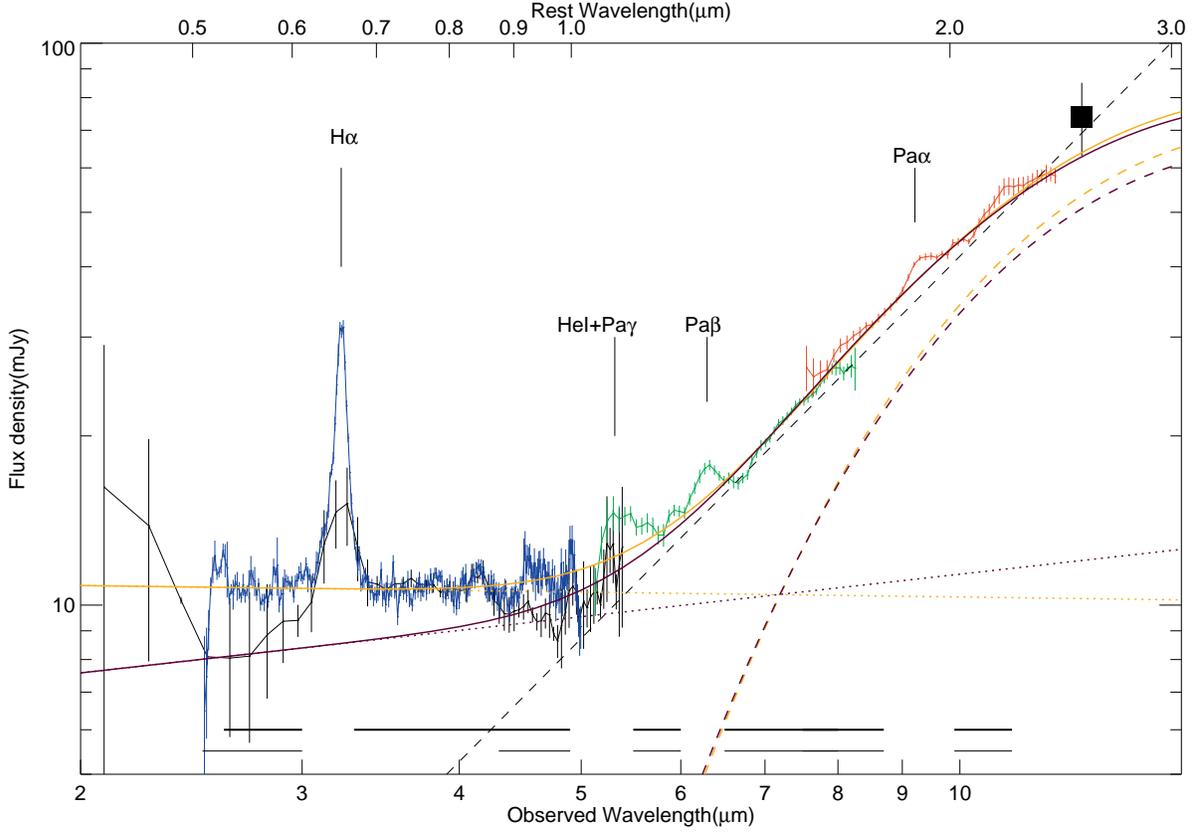}
    \caption{The NP (black line), NG (blue), SG1 (green) and SG2 (red)
      spectra of 
      APM 08279+5255, plotted as flux density 
      vs. wavelength (observed, bottom scale; rest-frame,
      top-scale). The vertical lines indicate the measurement
      error. A power-law fit, $f_{\nu} \sim \nu^{-2.26}$, to the
      continuum for $\lambda_{obs} < 15$ \micron, which is derived
      from the $Spitzer$ IRS observation, is shown as a black dashed
      line \citep{soifer04}. The filled square presents  the
      12.5\micron~flux density measured by \citet{egami00}.
      Two combined fits shown in orange and purple solid lines indicates
      two kinds of fits: an orange one used NG data for NIR data and
      a purple fitted a dip around 2.3-3\micron~on NP data corresponding
      to an extreme case.
      The
      power-law components are shown in orange and purple dotted lines, and the
      blackbodies shown in dashed lines with the same colors. The black
      lines at the bottom mark the places used in the continuum
      fitting. The upper thick ones are used for the case of NG and the others
      are for NP data with a dip.
    }
    \label{fig:fig1}
\end{figure*}

\section{Results}
\label{sec:res}

Figure \ref{fig:fig1} shows the spectra of APM 08279+5255 in the NP, NG,
SG1 and SG2 bands. The continuum and emission lines are clearly
detected at 2 - 13\micron. 

The continuum shows two components; 
one is a flat power-law at $\lambda_{\mathrm{rest}} 
\lesssim 1$\micron, whereas the other
is a power-law or bump sharply rising beyond 
$\lambda_{\mathrm{rest}} \gtrsim 1$\micron, where $\lambda_{\mathrm{rest}}$ 
denotes rest-frame wavelengths. 
Assuming the models consisting of two power-law components, the best fit to 
the data is $f_{\nu}
\sim \nu^{0.03}$ for $\lambda_{\mathrm{rest}} < 1$ \micron~and $f_{\nu}
\sim \nu^{-2.23}$ for $ \lambda_{\mathrm{rest}} > 1$ \micron.
The overall continuum level and the power-law slope for
$\lambda_{\mathrm{rest}} > 1\mu\mathrm{m}$ is consistent with observations of
$Spitzer$ IRS spectroscopy \citep{soifer04}.

\begin{figure*}
    \plottwo{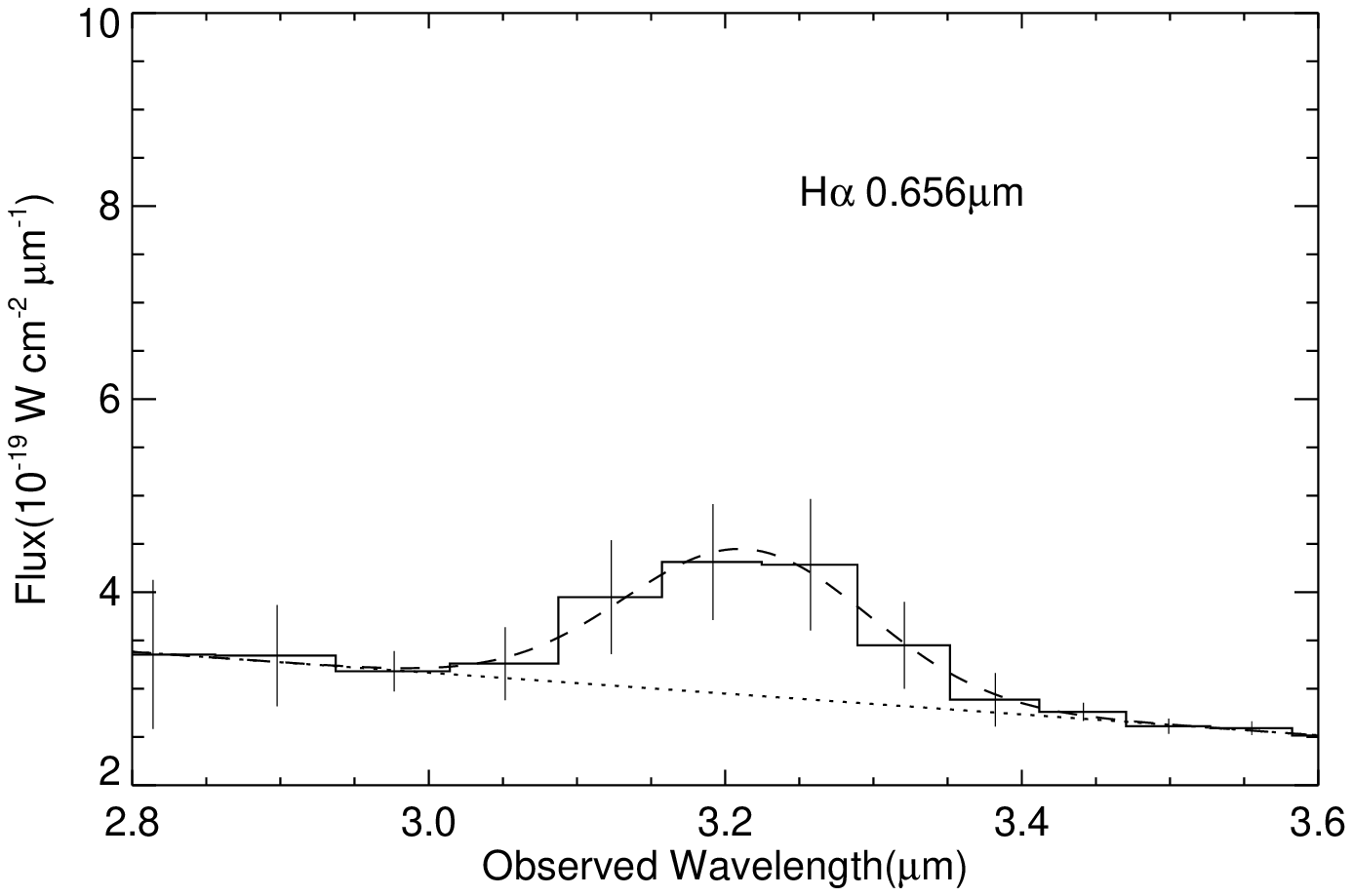}{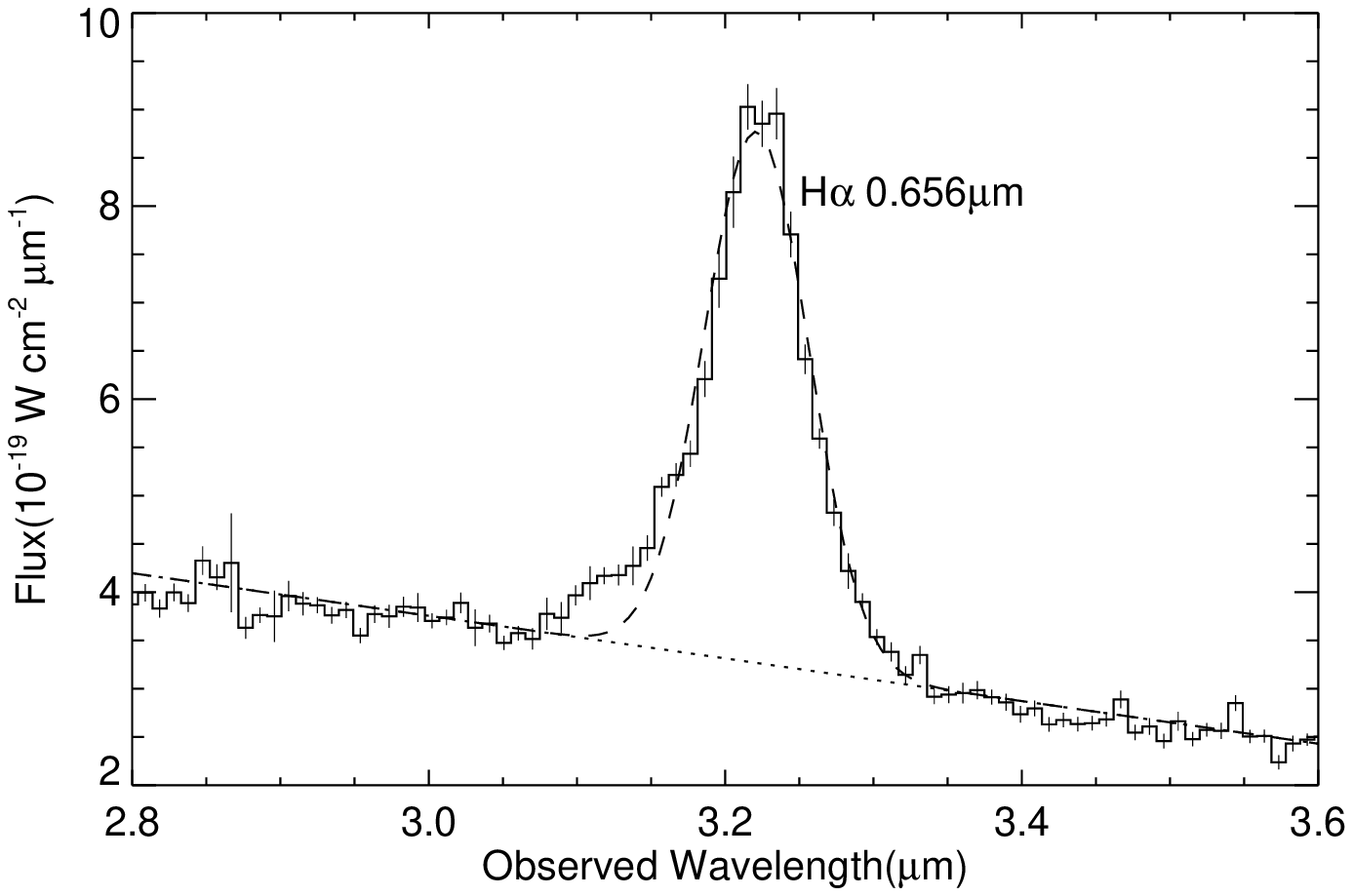}\\
    \plottwo{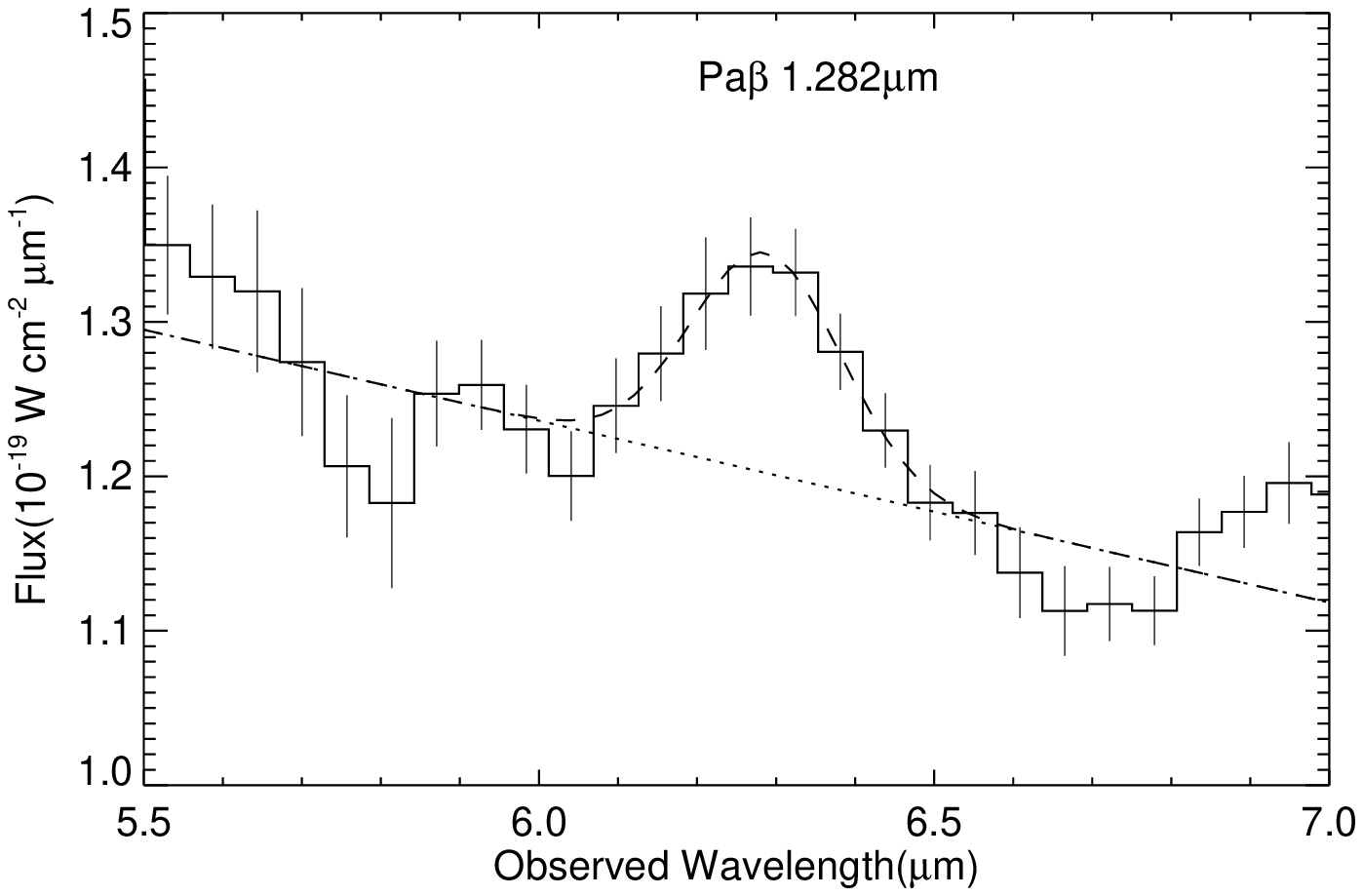}{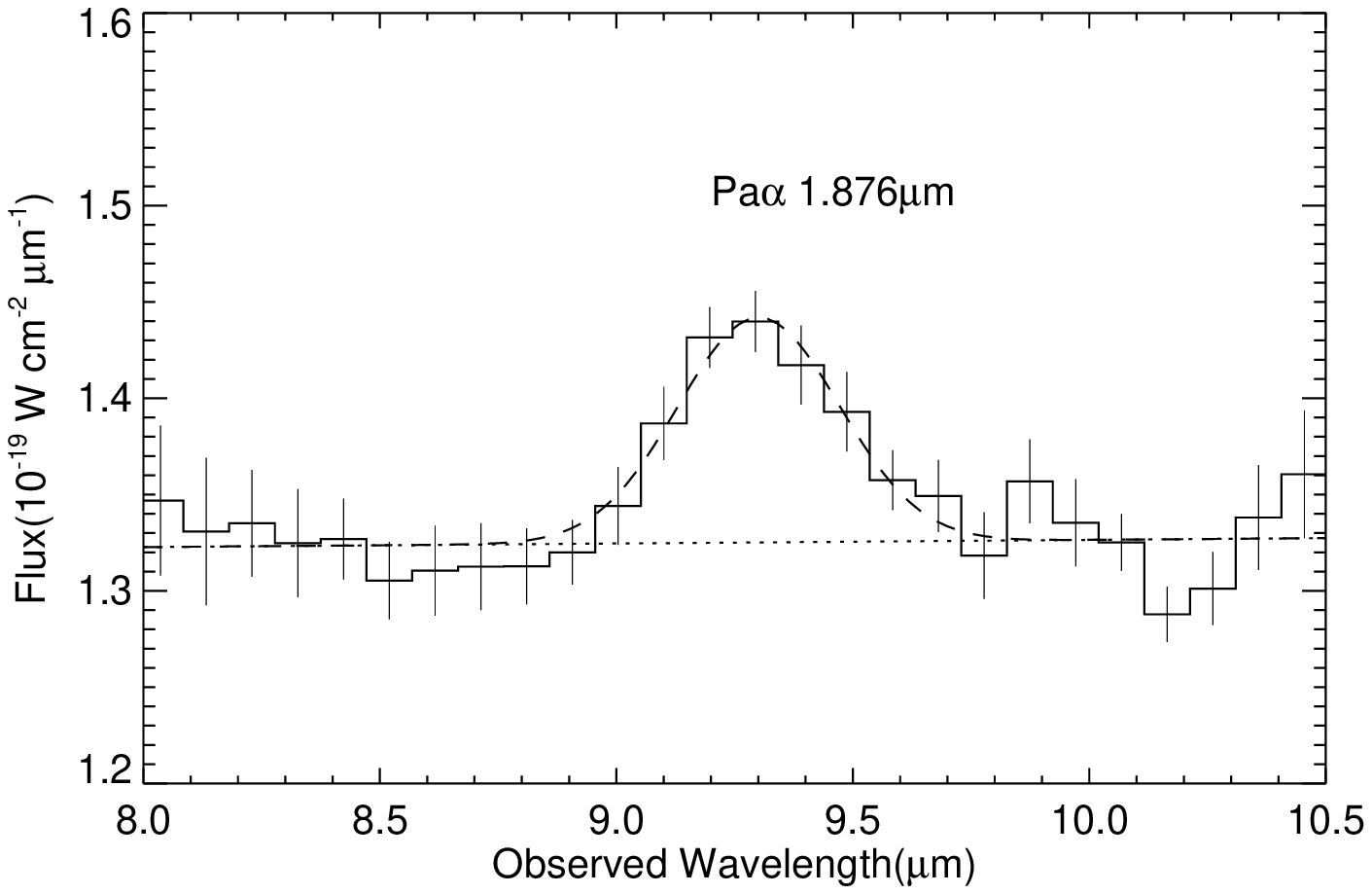}
    \caption{Profiles of emission lines: (Top-left) H$\alpha$ on NP,
      (Top-right) H$\alpha$ on NG, (Bottom-left) Pa$\beta$ on SG1, and
      (Bottom-right) Pa$\beta$ on SG2.
      The histograms are from $AKARI$ data. Vertical lines represents
      their errors. Dotted and 
    dashed lines show the fitted continuum and an emission line,
    respectively.}
    \label{fig:fig2}
\end{figure*}

HI recombination lines of H$\alpha$(0.656\micron), 
Pa$\alpha$(1.875\micron), and Pa$\beta$(1.282\micron) are clearly 
detected at 3.21, 6.3, and 9.3 \micron. The HeI $2^3S-2^3P$(1.083\micron) 
line and HI Pa$\gamma$(1.094\micron) line are blended and seen as a bump at
5.4\micron\footnote{
Models consisting of two lines poorly fit to 
the data; the OI $\lambda$11287\micron~emission line might contribute to 
the bump.}.
The line fluxes are measured, as illustrated in Figure \ref{fig:fig2}, 
by fitting a linear function to the local continuum data on either side 
of the line profile, and then fitting a Gaussian function to 
the linear-function-subtracted spectrum. 
The results including line fluxes and FWHMs are summarized 
in Table \ref{tab:result}. At the bottom of this table, the data obtained
by the $Spitzer$ \citep{soifer04} are given to facilitate the comparison.
The $Spitzer$ strengths of Pa$\alpha$ and Pa$\beta$ are significantly greater 
than those obtained with $AKARI$ by a factor of 1.5 - 2. This will be discussed 
later.

This is the first H$\alpha$ detection in APM 08279+5255. 
The H$\alpha$ fluxes in APM 08279+5255 are 60-100 times brighter than 
that in RX J1759.4+6638, a quasar at similar redshift, $z=4.3$ 
\citep{oyabu07}.
However, the line flux of H$\alpha$ is similar between 
them, consistent with the lens magnification of $\sim$100.

While the continuum fluxes observed on two occasions in a six-month interval 
are consistent with each other, the H$\alpha$ fluxes varied at a 2.3$\sigma$ 
level from the first occasion to the other. 
The H$\alpha$ flux measured with NP on the first occasion is
$F(\mathrm{H}\alpha)=(313\pm84)\times 10^{-22}\ \mathrm{W}\
\mathrm{cm}^{-2}$, while the second measurement with NG gives 
$F(\mathrm{H}\alpha)=(509\pm7)\times 10^{-22}\ \mathrm{W}\
\mathrm{cm}^{-2}$. Although the different dispersers were used, we believe 
that this variation is real> In fact, as compared in Table \ref{tab:result}, 
the Pa$\beta$ and Pa$\alpha$ strengths observed with $Spitzer$ 
differ by a factor of 1.5 - 2 from those observed with AKARI three year later. 
On the other hand, no significant variations in continuum have been 
observed between $Spitzer$ and $AKARI$ observations. This will be discussed 
again in Section \ref{sec:sec43}.

\begin{deluxetable}{llccccc}
    \tablewidth{16.5cm}
    \tabletypesize{\scriptsize}
    \tablecaption{Line flux measurements of APM 08279+5255
    \label{tab:result}}
    \tablehead{
      \colhead{Channel/} &
      \colhead{Line} & \colhead{Obs. Wavelength} &
      \colhead{Redshift} & \colhead{FWHM} &
      \colhead{Obs. Flux} & Obs. EW\tablenotemark{\star} \\
      Disperser & & ($\mu$m)& & (km\ s$^{-1}$) & ($10^{-22}\ \mathrm{W}\
      \mathrm{cm}^{-2}$) & ($\mu$m)}
    \startdata
    \multicolumn{7}{c}{Observation on 2006-10-19}\\
    \hline
    NIR/NP & H$\alpha$ 0.656\micron &  3.21 $\pm$ 0.04 &  3.90 $\pm$ 0.06 
    & $< 18300$ & 313 $\pm$ 84 & 0.107 $\pm$ 0.029
    \\
    MIR-S/SG1 & Pa$\beta$ 1.282\micron & 6.28 $\pm$ 0.03 & 3.90 $\pm$ 0.02
    & $< 12300$ & 42 $\pm$ 6 & 0.036 $\pm$ 0.005 \\
    MIR-S/SG2 & Pa$\alpha$ 1.875\micron & 9.30 $\pm$ 0.05 & 3.96 $\pm$ 0.02
    & $< 13800$ & 57 $\pm$ 6 & 0.043 $\pm$ 0.005 \\
    \hline
    \multicolumn{7}{c}{Observation on 2007-04-16}\\
    \hline
    NIR/NG & H$\alpha$ 0.656\micron &  3.22 $\pm$ 0.01 &  3.91 $\pm$ 0.02 &
    7721  & 509 $\pm$ 7 & 0.159 $\pm$ 0.002\\ 
    \hline
    \multicolumn{7}{c}{$Spitzer$ IRS Observation on 2003-11-23
      \citep{soifer04}}\\ 
    \hline
    & Pa$\beta$ 1.282\micron  & 6.315$\pm$0.013 & 3.925$\pm$0.006 &
    9500 & 81 $\pm$ 16 & 0.061 \\
    & Pa$\alpha$ 1.875\micron & 9.235$\pm$0.011 & 3.914$\pm$0.011 &
    8770 & 85 $\pm$ 11 & 0.059 \\
    \enddata
    \tablenotetext{\star}{The observed equivalent width (EW) of
      emission lines.}
\end{deluxetable}

\section{Discussion}
\label{sec:dis}

\subsection{Hot dust in a z=3.9 quasar}

The spectral energy distribution (SED) of quasars can be modelled with 
two components, big blue and infrared bumps \citep{sanders89,elvis94}. 
A big blue bump has been considered to be direct light from
the central engine, while the infrared bump is attributed to thermal emission 
by dust which absorbs UV/optical light and re-emit it in the infrared.
The sharp rise from $\lambda_{\mathrm{rest}} \sim 1$\micron\ toward 
the peak of the infrared bump is especially attributed to hot dust 
surrounding the central source. We assumed that the observed spectra only 
consist of a power-law emission and a single-temperature blackbody, 
representing the big blue bump and the infrared bump, respectively. 
These models are defined as:
\begin{equation}
    F_{\nu} = C_1 \nu^{\alpha}+C_2 B_{\nu}(T_{\mathrm{dust}})(1-\mathrm{e}^{-\tau_{\nu}}),
\end{equation}
where $\alpha$ is a power-law index, $B_{\nu}(T_{\mathrm{dust}})$
is the Planck function with dust temperature $T_{\mathrm{dust}}$, and
$\tau_{\nu}$ is the dust optical depth. For the frequency dependence of
the dust optical depth we adopt the result of \citet{wein01}. 

The parameters, $C_1,\alpha,C_2,T_{\mathrm{dust}}$ and $\tau_{\nu}$ are derived by
$\chi^2$ fitting to the observed continuum. Near-infrared
part of the spectrum was  
taken with NP and NG, resulting in two sets of the spectra, namely, 
NP-SG1-SG2 and NG-SG1-SG2. 
Fits to these two spectra resulted in the allowable temperature in 1
$\sigma$ error,
$1270\mathrm{K} < T_{\mathrm{dust}} < 1295\mathrm{K}$, with $\alpha =
-0.02\pm0.03$ for NP and $1250\mathrm{K} <
T_{\mathrm{dust}} < 1295\mathrm{K}$ with $\alpha = 
0.03\pm0.03$ for NG. In the both case, the dust optical depths were
just determined to be the lower limit,
$\tau_{2\mu\mathrm{m}} \gtrsim 2$, which means that the hot dust
emission of APM 08279+5255 is optically thick at a rest wavelength
$\lesssim 2\mu\mathrm{m}$. Our observations directly detected the
optically thick region of hot dust in the quasar. 
These fits were 
made to the continua after removing significant emission features from
the observed spectra. If the broad deep dip at 2.5\micron~  
of the NP spectrum is real continuum, as an extreme case to check the
dependency 
of the dust temperature on the power-law index,
$\alpha = -0.25, T_{\mathrm{dust}}=1290\mathrm{K}$ and
$\tau_{2\mu\mathrm{m}} \gtrsim 2$ are obtained.
Thus, the dust temperature is 1280 $\pm 20K$.

Temperatures $\sim 1300K$ of this quasar suggest that graphite and
silicate grains can survive as the hot dust component in this quasar
because sublimation temperatures for graphite and silicate
grains are $T\sim 1800K$ and $\sim 1500K$, respectively
\citep{salpeter77,huffman77}. 
We interpret the existence of such hot dust whose temperature is close to
sublimation temperatures of dust as the innermost dust
component heated by the strong radiation of the AGN central
engine. The dust closer to the central engine has higher temperatures
but cannot survive inside any critical radius at which it begins to
sublimate. 

As we fit the hot dust emission using a single-temperature blackbody
of $\sim$ 1300 K in Figure \ref{fig:fig1}, the luminosity of the hot
dust 
is calculated at $L_{\mathrm{hot}} = 2.1 \times 10^{15}\ m^{-1}\ 
\mathrm{L_{\sun}}$, where $m$ is the magnification factor of the
gravitational lensing. 
If it is
assumed the blackbody emission is due 
to heated dust grains, a measurement of the size of emitting region
can be determined from the relationship,
\begin{equation}
    F_{\nu}=m \frac{(1-e^{-\tau_{\nu}}) B_{\nu}(T_{\mathrm{dust}})}{(1+z)^3} \pi \left(\frac{R_{\mathrm{d}}}{D_{\mathrm{A}}}\right)^2.
\end{equation}
Here $F_{\nu}$ is the observed flux
density, $m$ is the magnification factor, $\tau_{\nu}$ is the optical
depth,
$D_A$ is the angular diameter distance and $R_{\mathrm{d}}$ is the radius of
the emitting region. On the condition of $T_{d} \sim 1300$K and
optical thickness, $\tau_{2\mu\mathrm{m}} \gtrsim 2$, as derived above,
the radius of emitting region is $R_{\mathrm{d}}
= 10\ m^{-1/2}\ \mathrm{pc}$. The size of the emitting region is
small enough  
to magnify the flux of 
hot dust $\sim 100$ times \citep{egami00}. 

To measure the dust mass, $M_{\mathrm{hot}}$, which emit in the
near-infrared, we used the 
following equation,
\begin{equation}
    M_{\mathrm{hot}}=\frac{\tau_\nu}{k_d}\ \pi\ R_{\mathrm{d}}^2,
\end{equation}
where $k_{d}$ is the mass absorption
coefficient according to 
\citet{wein01}. Now our fitting result is  the optically
thick condition, $\tau_{2\mu\mathrm{m}} > 2$, and thus we can obtain
the lower limit of the dust mass, $M_{\mathrm{hot}} > 20\
m^{-1}\ \mathrm{M_{\sun}}$ using 2\micron~flux in the rest frame. 
We emphasize that the hot dust mass is much less than $7.5 \times
10^8\ m^{-1}\ 
\mathrm{M_{\sun}}$ from warm dust with $T_{\mathrm{dust}} \sim 215
\mathrm{K}$ and $2.6 \times 10^9\ m^{-1}\ 
\mathrm{M_{\sun}}$ from cold one with $T_{\mathrm{dust}} \sim 65
\mathrm{K}$ in \citet{weiss07}, 
in which the continuum 
fluxes measured in the 
centimeter and millimeter are fit with the 2-component dust model and
their masses are calculated. 

As studies of hot dust in low-redshift quasars, \citet{kobayashi93}
observed 
low-resolution near-infrared spectra of 14
quasars with redshift $z<0.3$ and reproduced their
continua from 0.95 to
2.5\micron~with a combination of two radiation components; the
power-law and blackbody of 1470$\pm$90K. 
There is also a recent work that \citet{glikman06} constructed the
composite spectrum from 
0.58 to 3.5\micron~from near-infrared observations of 27 quasars
with the redshift of $0.118 < z <0.418$. Their composite spectrum is
fitted with a spectral slope of -0.92 and a blackbody temperature of
1260K. 
The dust temperature of APM 08279+5255 does not show a big difference
from those of
low-redshift quasars. 


\subsection{Hydrogen recombination lines}

Our $\mathrm{Pa}\alpha/\mathrm{Pa}\beta$ is $1.36 \pm 0.24$ in APM 08279+5255, 
which agrees with $1.05 \pm 0.25$ observed by \citet{soifer04}. \citet{landt08} 
reported $\mathrm{Pa}\alpha/\mathrm{Pa}\beta$ of $1.27\pm0.07$ for 16 low-redshift quasars. It should be noted that 
$\mathrm{Pa}\alpha/\mathrm{Pa}\beta$ of $0.64 
\pm 0.01$ was obtained from a composite spectrum of low-redshift quasars
\citep{glikman06}. In any case, $\mathrm{Pa}\alpha/\mathrm{Pa}\beta$ observed in quasars is significantly smaller than 
2.0 predicted from Case B recombination \citep{osterbrock89},
suggesting HI recombination lines come from BLR clouds where the gas
density is very high. 

\begin{figure*}
    \plotone{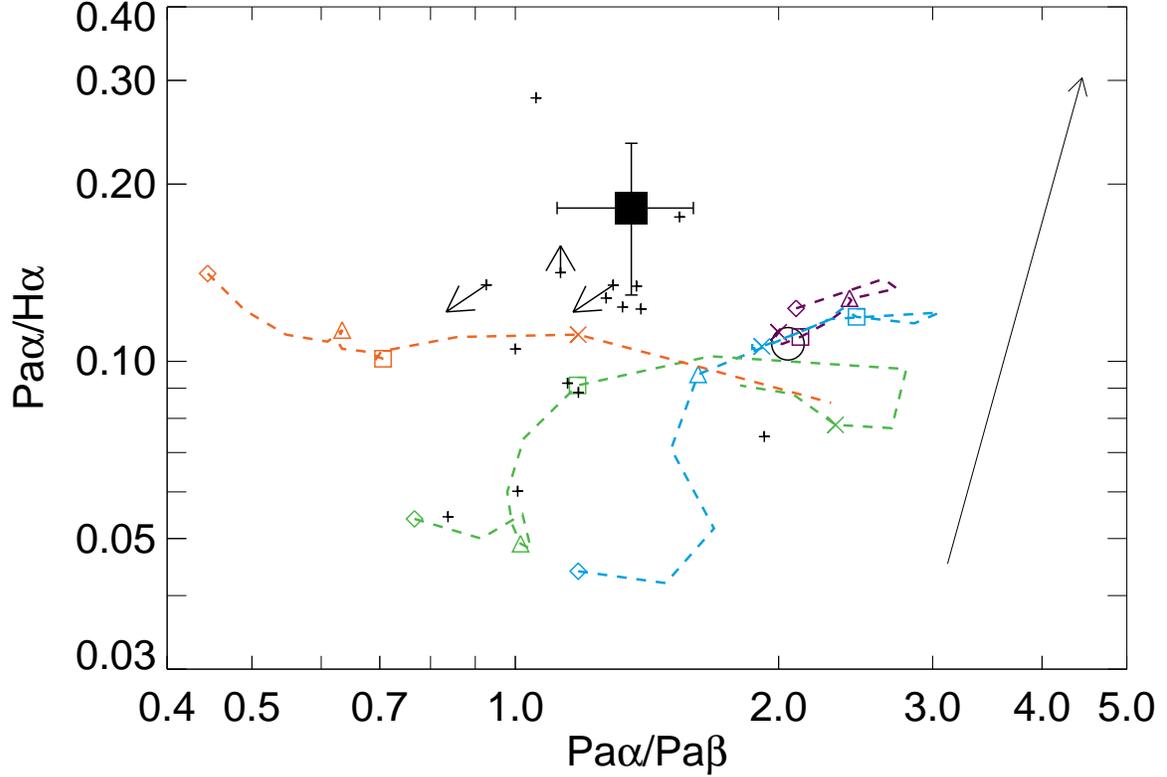}
    \caption{The $\mathrm{Pa}\alpha/\mathrm{Pa}\beta$ line flux ratio
      vs. the 
      $\mathrm{Pa}\alpha/\mathrm{H}\alpha$ ratio of APM 08279+5255
      (large filled square calculated with H$\alpha$ flux for NP). A open circle represents Case-B
      values with $T=10^4\mathrm{K}$ and $n_{\mathrm{H}}=10^4\
      \mathrm{cm}^{-3}$ \citep{osterbrock89}. Small crosses represent the values of
      low-redshift quasars \citep{landt08}. Red, green, blue and
      purple dashed lines show $\log n_{\mathrm{H}} =
      14,12,10,$ and 8, respectively. 
  The diamond, triangle, square and
      cross marks show $\log U=0,-2,-4$ and $-6$. 
Details of the models is in
      text. An arrow
      extends out along the reddening line to $E(B-V)=1$.      
    }
    \label{fig:fig4}
\end{figure*}

Figure \ref{fig:fig4} plots $\mathrm{Pa}\alpha/\mathrm{Pa}\beta$
versus $\mathrm{Pa}\alpha/\mathrm{H}\alpha$ of APM 08279+5255 along with 
low-redshift quasars\citep{landt08}. The APM 08279+5255 data used 
in this Figure were taken in 2006 October when the three lines were observed
simultaneously. Photoionization codes Cloudy(version 06.02 by \citet{ferland98})was used to probe BLR clouds. Our 
models for BLR clouds and the incident 
continuum from the central sources are the same as 
as those used in \citet{tsuzuki06} and \citet{matsuoka07,matsuoka08}. 
The incident continuum is defined as:
\begin{equation}
f_{\nu} \propto
\nu^{\alpha_{\mathrm{UV}}} \exp(-h\nu/kT_{\mathrm{CUT}})
\exp(-kT_{\mathrm{IR}}/h\nu) + a \nu^{\alpha_{\mathrm{X}}}, 
\end{equation}
where
$T_{\mathrm{CUT}}$ and $T_{\mathrm{IR}}$ is the high-energy and
low-energy cut-off temperature, respectively.
The low-energy cutoff temperature has little effect on the spectral result,
thus a temperature $kT_{\mathrm{IR}}=0.01\mathrm{Ryd}$ is preset for the UV
bump cutoff in the infrared.
The UV and X-ray continuum components are combined
using a UV to X-ray logarithmic spectra slope $\alpha_{OX}$. 
We adopted a parameter set of 
$[T_{\mathrm{CUT}},\alpha_{\mathrm{UV}},\alpha_{\mathrm{X}},\alpha_{\mathrm{OX}}=1.5\times
10^5 \mathrm{K},-0.2,-1 .8,-1.4]$. The gas was modeled to have a
constant hydrogen density $n_H$ and exposed to the ionizing continuum
radiation expressed with an ionization parameter U. 
We performed the
calculation with the $(n_H,U)$ sets in a range of $10^{8}\
\mathrm{cm}^{-3} \leq n_{H} \leq 10^{14}\ \mathrm{cm}^{-3}$ and
$10^{-7} \leq U \leq 10^0$ stepped by 0.5 dex. 

As can see in Figure \ref{fig:fig4}, none of quasars agree with Case B
recombination, regardless of intrinsic reddening to BLR clouds. 
The APM 08279+5255 data are reasonably represented with 
a parameter set of $\log n_{H} \sim 12- 14$ and $\log U \sim -2 - -6$ with
$E(B-V)\sim 0.3-0.6$. 
Though optically thick region of hot dust surrounds the BLR, the
intrinsic extinction in the BLR is moderate. This supports an idea of
non-spherical distribution of hot dust near the central source, which is
consistent with the AGN model with the dust torus \citep{anto85}. 

\subsection{Variability of HI recombination lines}
\label{sec:sec43}

As discussed in Section \ref{sec:res}, the line fluxes of Pa$\alpha$ and 
Pa$\beta$ have decreased by a factor of 1.5 - 2 in three years. In addition, 
the H$\alpha$ flux also varied in six months. 
On the other hand, our continuum level of dust emission
in $\lambda_{rest} \gtrsim 1$\micron~was kept constant over these periods.
This means that the brightness in the BLR varied while that in thermal emission from hot dust was constant. 

The variability of the broad lines and the constancy of the dust continuum can
be explained in terms of the source size of the BLR and the hot-dust emitting 
regions. 
For example, the size of the BLR of NGC 4151 is a few to 10 light days
\citep[i.e.][]{clavel90,maoz91,kaspi96}, while 
the inner boundary of the dusty torus extends to 48 
light days from the central source \citep{minezaki04}.
In the central source, the gas in the accretion disk, which is falling to 
the supermassive black hole, radiates the continuum from the high-energy to 
the optical. Part of this continuum is absorbed in BLR clouds and 
thus ionized gas produces line emission such as HI recombination lines. 

The incident continuum is also absorbed by dust which mainly exist at a 
distance greater than BLR clouds from the center.
When the brightness of the incident continuum varies, BLR lines vary 
in a time-delay of days or months and the thermal emission in the near- and 
mid-infrared radiation also varies after the time-delay of months or years.
Thus, the variation in the near- and mid-infrared radiation is smeared out more than that in recombination lines in 
BLR gas, making more difficult to detect 
variations in emission from hot dust than recombination lines.

It is unlikely that the variation in the HI recombination lines is
attributed to microlensing events which have been 
observed lensed quasars
\citep{burud00,burud02,hjorth02,jakobsson05}. The lensing galaxy of
APM 08279+5255 has not 
been 
found. If the lensing galaxy is at $z=1-3$, the Einstein radius of a star 
with $1 \mathrm{M}_{\sun}$ in the lensing galaxy is 
$\theta_{\mathrm{E}} \sim 1 \times
10^{-6}\ \mathrm{arcsec}$, or $7 \times 10^{-3}\ \mathrm{pc}$ 
at $z=3.91$. 
This is much smaller than the size of hot dust which emits in the
infrared, $10\ m^{-1/2}\ \mathrm{pc}$. Therefore, the microlensing
event of a star in the lensing galaxy is unable to magnify infrared
emission, while it is able to magnify fluxes coming from smaller
regions like BLR clouds or an accretion disk.
However, assuming the star in the lensing galaxy having a velocity of
$300\ \mathrm{km}\ \mathrm{s}^{-1}$, 
the timescale of $\sim 30-100\ \mathrm{year}$ is longer than
that of our detection of the variability \citep{chiba05}.
Therefore, it might be difficult to explain that the
variability is caused by microlensing. 

\section{Summary}
\label{sec:sum}

This paper presents near- and mid-infrared spectra of
a gravitationally lensed quasar APM 08279+5255 at $z=3.91$ with
AKARI/IRC. The observations were performed with the IRC onboard the
$AKARI$ infrared satellite. We have detected the continuum from 2\micron~to
13\micron~in the observed frame, corresponding to 0.5\micron~to
2.4\micron~ in the rest-frame, and hydrogen emission lines of
H$\alpha$, Pa$\beta$ and Pa$\alpha$ and probably HeI emission
line. The major conclusions are as follows. 
\begin{enumerate}
\item The thermal emission of hot dust heated by the central engine of this
    quasar is measured by the spectral fitting with the simple model
    expressed with a power-law and a blackbody.
    The thermal emission represents optically thick emission by hot
    dust at $T\sim 1300\mathrm{K}$ with $\tau_{2\mu\mathrm{m}} > 2$ and
    mass, $M_{\mathrm{hot}} > 10\ m^{-1/2}\
    \mathrm{M}_{\sun}$. The temperature is consistent with those of
    low-redshift quasars which were also measured
    spectroscopically\citep{kobayashi93,glikman06}.
\item The flux ratios of Hydrogen emission lines, H$\alpha$, Pa$\beta$
    and Pa$\alpha$, can not be explained with the simple Case B
    recombination model, but they are consistent with those of low-redshift
    sample \citep{landt08}. Compared with the photoionization model,
    the physical condition of BLR in this quasar have a parameter set
    of $\log n_{H} \sim 12- 14$ and $\log U \sim -2 - -6$ with $E(B-V)\sim
    0.3-0.6$. The moderate extinction in the BLR and optically thick
    emission of hot dust support an idea of non-spherical distribution
    of dust near the central source, consistent with the AGN model
    with the dust torus.
\item The difference between $AKARI$ and $Spitzer$ measurements in
    Pa$\beta$ and Pa$\alpha$ emission lines could be explained by the
    variability of this quasar. The variability of the H$\alpha$ emission
    line was also detected with multiple observations with $AKARI$ in
    a span of six months. There are two possibilities of variability
    for this quasars: the intrinsic variability of this quasar itself and the
    microlensing caused by a star in a lensing galaxy. 
    Considering the timescale of microlensing,
    it might be ruled out for an explanation of the variability.
\end{enumerate}

Our $AKARI$ spectroscopic observations revealed that APM 08279+5255 at
$z=3.91$ show a lack of 
evolution in 
the BLR and the hot dust component. This suggests that these
components in this quasar have reached 
maturity very early on. 
This conclusion is probably applicable to most of quasars because 
the studies of broad emission lines in high-redshift quasars show the
lack of the evolution \citep[i.e.][]{iwamuro04}
and the infrared photometric studies
\citep{hines06, jiang06} of
quasars in $4.5 < z < 6.4$ do
not differ significantly from those in low-redshift.

\acknowledgments

$AKARI$ is a JAXA project with the participation of ESA. We thank all
the members of the $AKARI$ project for their continuous help and
support.


\begin{thebibliography}{}
\bibitem[Antonucci 
\& Miller(1985)]{anto85} Antonucci, R.~R.~J., \& Miller, J.~S.\ 1985, \apj, 297, 621
\bibitem[Barth et al.(2003)]{barth03} Barth, A.~J., Martini, 
P., Nelson, C.~H., \& Ho, L.~C.\ 2003, \apjl, 594, L95 
\bibitem[Boyle et al.(2000)]{boyle00} Boyle, B.~J., Shanks, T., 
Croom, S.~M., Smith, R.~J., Miller, L., Loaring, N., 
\& Heymans, C.\ 2000, \mnras, 317, 1014 
\bibitem[Burud et al.(2000)]{burud00} Burud, I., et al.\ 2000, 
\apj, 544, 117 
\bibitem[Burud et al.(2002)]{burud02} Burud, I., et al.\ 2002, \aap, 391, 481 
\bibitem[Chiba et al.(2005)]{chiba05} Chiba, M., Minezaki, T., 
Kashikawa, N., Kataza, H., \& Inoue, K.~T.\ 2005, \apj, 627, 53 
\bibitem[Clavel et al.(1990)]{clavel90} Clavel, J., et al.\ 
1990, \mnras, 246, 668 
\bibitem[Dietrich et al.(2003)]{dietrich03} Dietrich, M., Hamann, 
F., Shields, J.~C., Constantin, A., Heidt, J., J{\"a}ger, K., Vestergaard, 
M., \& Wagner, S.~J.\ 2003, \apj, 589, 722 
\bibitem[Downes et al.(1999)]{downes99} Downes, D., Neri, R., Wiklind, T.,
    Wilner, D.J., \& Shaver, P.A., 1999, \apj, 513, L1
\bibitem[Egami et al.(2000)]{egami00} Egami, E., Neugebauer, G., Soifer,
    B. T., Matthews, K., Ressler, M., Becklin, E. E., Murphy, T. W.,
    \& Dale, D. A. 2000, ApJ, 535, 561 
\bibitem[Elvis et al.(1994)]{elvis94} Elvis, M., et al.\ 1994, 
\apjs, 95, 1
\bibitem[Ferland et al.(1998)]{ferland98} Ferland, G.~J., 
Korista, K.~T., Verner, D.~A., Ferguson, J.~W., Kingdon, J.~B., 
\& Verner, E.~M.\ 1998, \pasp, 110, 761 
\bibitem[Glikman et al.(2006)]{glikman06} Glikman, E., Helfand, 
D.~J., \& White, R.~L.\ 2006, \apj, 640, 579 
\bibitem[Nagao et al.(2006)]{nagao06} Nagao, T., Marconi, A., \& Maiolino, R.\ 2006, \aap, 447, 157 
\bibitem[Hines et al.(2006)]{hines06} Hines, D.~C., Krause, O., 
Rieke, G.~H., Fan, X., Blaylock, M., 
\& Neugebauer, G.\ 2006, \apjl, 641, L85 
\bibitem[Hjorth et al.(2002)]{hjorth02} Hjorth, J., et al.\ 
2002, \apjl, 572, L11 
\bibitem[Huffman(1977)]{huffman77} Huffman, D.~R.\ 1977, Advances 
in Physics, 26, 129 
\bibitem[Ibata et al.(1999)]{ibata99} Ibata, R.~A., Lewis, 
G.~F., Irwin, M.~J., Leh{\'a}r, J., \& Totten, E.~J.\ 1999, \aj, 118, 1922 
\bibitem[Irwin et al.(1998)]{irwin98}Irwin, M.J., Ibata, R.A., Lewis, G.F.,
    \& Totten, E.J. 1998, \apj, 505, 529
\bibitem[Iwamuro et al.(2004)]{iwamuro04} Iwamuro, F., Kimura, 
M., Eto, S., Maihara, T., Motohara, K., Yoshii, Y., \& Doi, M.\ 2004, \apj, 614, 69 
\bibitem[Jakobsson et  al.(2005)]{jakobsson05} Jakobsson, P., Hjorth, J., Burud, I., Letawe, G., Lidman, C., \& 
Courbin, F.\ 2005, \aap, 431, 103 
\bibitem[Jiang et al.(2006)]{jiang06} Jiang, L., et al.\ 2006, 
\aj, 132, 2127 
\bibitem[Kaspi et al.(1996)]{kaspi96} Kaspi, S., et al.\ 1996, 
\apj, 470, 336 
\bibitem[Kawara et al.(1996)]{kawara96} Kawara, K., Murayama, 
T., Taniguchi, Y., \& Arimoto, N.\ 1996, \apjl, 470, L85 
\bibitem[Kawada et al.(2007)]{kawada07} Kawada, M., et al.\ 
2007, \pasj, 59, 389 
\bibitem[Kobayashi et al.(1993)]{kobayashi93} Kobayashi, Y., Sato,
    S., Yamashita, T, Shiba, H, and Takami, H.\ 1993, \apj, 404, 94
\bibitem[Kurk et al.(2007)]{kurk07} Kurk, J.~D., et al.\ 2007, 
\apj, 669, 32 
\bibitem[Landt et al.(2008)]{landt08} Landt, H., Bentz, M.~C., 
Ward, M.~J., Elvis, M., Peterson, B.~M., Korista, K.~T., 
\& Karovska, M.\ 2008, \apjs, 174, 282 
\bibitem[Maoz et al.(1991)]{maoz91} Maoz, D., et al.\ 1991, 
\apj, 367, 493 
\bibitem[Matsuoka et al.(2007)]{matsuoka07} Matsuoka, Y., Oyabu, 
S., Tsuzuki, Y., \& Kawara, K.\ 2007, \apj, 663, 781 
\bibitem[Matsuoka et al.(2008)]{matsuoka08} Matsuoka, Y., Kawara, 
K., \& Oyabu, S.\ 2008, \apj, 673, 62 
\bibitem[Minezaki et al.(2004)]{minezaki04} Minezaki, T., Yoshii, 
Y., Kobayashi, Y., Enya, K., Suganuma, M., Tomita, H., Aoki, T., 
\& Peterson, B.~A.\ 2004, \apjl, 600, L35 
\bibitem[Moshir et al.(1992)]{moshir92} Moshir, M., Kopman, G., 
\& Conrow, T.~A.~O.\ 1992, Pasadena: Infrared Processing and Analysis Center, California Institute of Technology, 
1992, edited by Moshir, M.; Kopman, G.; Conrow, T.~a.o.,  
\bibitem[Murakami et al.(2007)]{murakami07} Murakami, H., et al.\ 
2007, \pasj, 59, 369 
\bibitem[Ohyama et al.(2007)]{ohyama07} Ohyama, Y., et al.\ 
2007, \pasj, 59, 411
\bibitem[Onaka et al.(2007)]{onaka07} Onaka, T., et al.\ 2007, 
\pasj, 59, 401 
\bibitem[Osterbrock (1989)]{osterbrock89} Osterbrock, D. E., 1989,
    Astrophysics of Gaseous Nebulae and Active Galactic Nuclei
    (University Science Books)
\bibitem[Oyabu et 
al.(2001)]{oyabu01} Oyabu, S., et al.\ 2001, \aap, 365, 409 
\bibitem[Oyabu et al.(2007)]{oyabu07} Oyabu, S., et al.\ 2007, 
\pasj, 59, 497 
\bibitem[Salpeter(1977)]{salpeter77} Salpeter, E.~E.\ 1977, \araa, 15, 267 
\bibitem[Sanders et al.(1989)]{sanders89} Sanders, D.~B., 
Phinney, E.~S., Neugebauer, G., Soifer, B.~T., 
\& Matthews, K.\ 1989, \apj, 347, 29 
\bibitem[Soifer et al.(2004)]{soifer04} Soifer, B.T., Charmandaris,
        V., Brandl, B.R. et al. 2004, ApJS, 154, 151
\bibitem[Tsuzuki et al.(2006)]{tsuzuki06} Tsuzuki, Y., Kawara, 
K., Yoshii, Y., Oyabu, S., Tanab{\'e}, T., 
\& Matsuoka, Y.\ 2006, \apj, 650, 57 
\bibitem[Wei{\ss} et 
al.(2007)]{weiss07} Wei{\ss}, A., Downes, D., Neri, R.,
Walter, F., Henkel, C., Wilner, D.~J., Wagg, J., \& Wiklind, T.\ 2007,
\aap, 467, 955  
\bibitem[Weingartner 
\& Draine(2001)]{wein01} Weingartner, J.~C., \& Draine, B.~T.\ 2001, \apj, 548, 296 
\bibitem[Wills et al.(1985)]{wills85} Wills, B.~J., Netzer, H., 
\& Wills, D.\ 1985, \apj, 288, 94 
\bibitem[York et al.(2000)]{york00} York, D.~G., et al.\ 2000, 
\aj, 120, 1579 
\bibitem[Yoshii et al.(1998)]{yoshii98} Yoshii, Y., Tsujimoto, 
T., \& Kawara, K.\ 1998, \apjl, 507, L113 
\end{thebibliography}
\end{document}